\documentclass[prb,floatfix,twocolumn,showpacs,amsmath,amssymb,
superscriptaddress]{revtex4}
\usepackage{graphicx}
\usepackage{dcolumn}
\usepackage{bm}
\begin{document}

\title{Theoretical constraints for the magnetic-dimer transition in 
two-dimensional spin models}
\author{Leonardo Spanu}
\affiliation{INFM-Democritos, National Simulation Center and International
School for Advanced Studies (SISSA), I-34014 Trieste, Italy}
\author{Federico Becca}
\affiliation{INFM-Democritos, National Simulation Center and International
School for Advanced Studies (SISSA), I-34014 Trieste, Italy}
\affiliation{Laboratoire de Physique Th\'eorique, Universit\'e ``Paul Sabatier'', F-31062 Toulouse, France}
\author{Sandro Sorella}
\affiliation{INFM-Democritos, National Simulation Center and International
School for Advanced Studies (SISSA), I-34014 Trieste, Italy}
\date{\today}

\begin{abstract}
From general arguments, that are valid for spin models with sufficiently
short-range interactions, we derive strong constraints on the excitation
spectrum across a continuous phase transition at zero temperature between a 
magnetic and a dimerized phase, that breaks the translational symmetry. 
From the different symmetries of the two phases, it is possible to predict,
at the quantum critical point, a branch of gapless excitations, not 
described by standard semi-classical approaches. By using these arguments, 
supported by intensive numerical calculations, we obtain a rather convincing 
evidence in favor of a first-order transition from the ferromagnetic to the
dimerized phase in the two-dimensional spin-half 
model with four-spin ring-exchange interaction, recently introduced by 
A.W. Sandvik {\it et al.} [Phys. Rev. Lett. {\bf 89}, 247201 (2002)]. 
\end{abstract}

\maketitle

\section{Introduction}

In the last years, a large amount of work has been done to clarify the 
properties of unconventional magnetic systems. Indeed, the presence of 
frustrated interactions, generated either by the geometry 
of the lattice or by the competing interactions, can give rise 
to many anomalous low-energy properties.~\cite{diep} One of the most 
exciting is the possibility to obtain phases with no magnetic order even
at zero temperature.~\cite{fazekas}
The actual interest in this subject has been recently renewed by the discovery 
of several materials~\cite{coldea,kanoda,kagome} that do not show any sign 
of magnetic order down to very low temperatures.
From a theoretical point of view, a considerable progress has been done in 
elucidating the possible unconventional states and, apart from magnetic 
phases, it is now well accepted that many systems exhibit a large variety 
of disordered ground states. 
One class of such paramagnetic states is given by the so-called valence-bond 
solids, where pairs of nearest-neighbor spins form a singlet, leading to an
ordered pattern of valence bonds. These states explicitly break the 
translational symmetry, implying a degenerate ground state, and represent the
two-dimensional extension of the celebrated dimerized state, rather well 
established in quasi-one-dimensional materials.~\cite{oned}
Another class of disordered phases is given by the so-called spin liquids,
that do not break any local symmetry. It is now rather widely accepted 
that also spin liquids possess some degeneracy, or even a gapless spectrum,
and have some kind of order, called topological order, related to non-local 
operators.~\cite{wen,hastings}

Much recent research has focused on the way to describe transitions between 
different quantum phases. Indeed, from the Landau's arguments valid for 
classical critical phenomena, if one exclude a delicate fine tuning of the 
parameters, there is no reason to have a continuous transition between two 
phases with different symmetries of the order parameters, like for instance
between a magnetic system and a valence-bond solid, and a first-order 
transition would be the most natural scenario.
However, very recently,~\cite{senthil1,senthil2} these arguments have 
been questioned and it has been argued that the situation can be more 
complex and richer in quantum systems, where the Landau paradigm could be 
violated, giving rise to an appealing way to escape a first-order 
transition. Therefore, it has been suggested that a quantum phase transition 
is intrinsically different from a classical one. In particular, when 
considering quantum phases, the natural description is no longer by using
order parameters but instead by considering new degrees of freedom
carrying fractional quantum numbers, that emerge at the critical point.

The arguments given in Refs.~\onlinecite{senthil1} and~\onlinecite{senthil2},
based on field-theory approaches, are rather convincing but it is natural to 
ask if microscopic spin models, defined on a lattice, could present the same 
low-energy critical behavior.
In this sense, however, we face a technical difficulty: a good diagnosis 
would require to study large enough clusters, but this is not possible since
quantum Monte Carlo simulations of frustrated antiferromagnets are plagued 
with a very severe minus sign problem.
In this respect, it is important to find out microscopic models that do not 
suffer from this problem but, at the same time, present phase transitions 
between quantum phases. Recently, Sandvik and collaborators~\cite{sandvik} 
introduced a spin-half model on the two-dimensional square lattice in the 
presence of a particular ring-exchange interaction:
\begin{equation}\label{hamiring}
{\cal H} = {\cal H}_{\rm XY} + {\cal H}_{\rm Ring},
\end{equation}
where
\begin{eqnarray}
{\cal H}_{\rm XY} &=& -\frac{J}{2} \sum_{\langle i,j \rangle} (S^+_i S^-_j +S^-_i S^+_j) \nonumber \\
{\cal H}_{\rm Ring} &=& -K \sum_{[i,j,k,l]} (S^+_i S^-_j S^+_k S^-_l + S^-_i S^+_j S^-_k S^+_l). \nonumber
\end{eqnarray}
Here $\langle i,j \rangle$ denotes a pair of nearest neighbor sites and 
$[i,j,k,l]$ indicates the sites on the corners of a plaquette. In particular,
${\cal H}_{\rm Ring}$ is a ring-exchange interactions that acts as a 
cyclic permutation on the spins of a given plaquette whenever they are in 
a $S^z=0$ configuration with parallel spins on opposite corners.~\cite{balents}
The case with $J>0$ and $K>0$ is considered, in order to avoid the sign 
problem. Notice that ${\cal H}$ is invariant under global U(1) 
transformations $S^+_i \to e^{i \theta} S^+_i$.

The Hamiltonian~(\ref{hamiring}) has been investigated by using quantum
Monte Carlo~\cite{sandvik,sandvik2} and it has been shown that, at zero 
temperature and zero magnetic field, there is a transition between a 
ferromagnetic phase, stable for small values of $K/J$, and a dimerized one, 
characterized by columnar order at $Q=(\pi,0)$ and $Q=(0,\pi)$, stable for 
intermediate ring-exchange couplings. Finally, for large $K/J$ a checkerboard 
phase is also expected.~\cite{sandvik,sandvik2}
Note that, in our notation, $J$ is twice time bigger than the one used in 
Refs.~\onlinecite{sandvik} and~\onlinecite{sandvik2}. Once scaled in our 
units, their best estimate 
for the transition point from the magnetic to the dimerized phase is 
$K_c/J \sim 3.957$.~\cite{thanks}
In the original paper,~\cite{sandvik} it has been shown that, by increasing 
the strength of the ring exchange, both the spin stiffness $\rho_s$ and the 
magnetization $M$ decrease and eventually vanish, with a corresponding 
insurgence of a finite dimer order parameter. Hence, the 
Hamiltonian~(\ref{hamiring}) has been suggested as a prototype model to 
test the ideas of Refs.~\onlinecite{senthil1} and~\onlinecite{senthil2}. 
However, subsequent and more accurate calculations have raised some doubt on 
the previous ones, suggesting instead a weak first-order 
transition.~\cite{note2} The actual nature of the transition is therefore
still controversial.

In this paper, we propose a different point of view to 
investigate the properties of a phase transition between a magnetic and a
dimerized phase of a generic spin Hamiltonian with sufficiently short-range
interactions. 
Indeed, instead of calculating the order parameters and the stiffness, we 
derive strong constraints on the excitation spectrum across a continuous 
phase transition between a magnetically ordered and a dimerized phase,
breaking translational symmetry. 
These constraints can be verified by using numerical techniques, such as the
Green's function Monte Carlo method.~\cite{gfmc}
In our opinion this way to proceed represents the first attempt to clarify 
the evolution of the excitation spectrum and can give important insight into 
the nature of the transition point also in other contexts.

The paper is organized as follow: in Sec.~\ref{sec1} we develop the 
general theoretical formalism, in Sec.~\ref{sec2} we present our numerical 
results, and in Sec.~\ref{sec3} we draw our conclusions.
 
\section{Theoretical formalism}\label{sec1}

In this section, we describe in detail the implications of a continuous
phase transition from a magnetically ordered phase to a dimerized one
on the excitation spectrum near the quantum critical point.
In the following, we will consider an Hamiltonian with a generic translational 
invariant two-spin coupling and SU(2) symmetry, and at the end of the section
we will also discuss the cases with U(1) symmetry or multi-spin interactions.

Let us consider the Heisenberg Hamiltonian on a two-dimensional lattice with
$N=L \times L$ sites:
\begin{equation}
{\cal H} = \sum_{i,j} J_{i,j} {\vec S}_i \cdot {\vec S}_j,
\end{equation}
where $J_{i,j}$ is the coupling between the spins at site $R_i$ and $R_j$ and
${\vec S}_i = (S_i^x, S_i^y, S_i^z)$ is the spin operator at the site $R_i$.
For a translational invariant magnetic coupling that depends only upon 
$|R_i-R_j|$, i.e., $J_{i,j}=J_{|i-j|}=J_r$, we can perform the Fourier 
transform and easily obtain:
\begin{equation}\label{ham}
{\cal H} = \sum_q J_q {\vec S}_{q} \cdot {\vec S}_{-q},
\end{equation}
where ${\vec S}_{q} = 1/\sqrt{N} \sum_j e^{i q R_j} {\vec S}_j$ and 
$J_q = \sum_{r} e^{i q R_r} J_{r}$ define the Fourier transform of the local
spin operator and the magnetic coupling, respectively.
In the following, we will consider interactions that are sufficiently 
short range, that is:
\begin{equation}\label{shortrange}
\tilde{J}=\frac{1}{2N} \sum_{i,j} |J_{i,j}| |R_i-R_j|^2 < \infty.
\end{equation}
For the SU(2) case, the eigenstates can be classified according to the total 
spin $S$ and the ground state, here denoted by $|\Psi_{Q_0} \rangle$,
is generally a singlet with momentum $Q_0$.
By means of the so-called Feynman construction, originally introduced to 
describe the excitations of the liquid Helium,~\cite{feynman} we can easily
define a variational state for the triplet excited state with momentum $q$ 
with respect to the ground state:
\begin{equation}
|\Psi_{Q_0+q} \rangle = S_q^\alpha|\Psi_{Q_0} \rangle,
\end{equation}
$S_q^\alpha$ being one of the three components of the total spin
${\vec S}_{q}$ ($\alpha = x,y,z$).
Then, it is easy to calculate the variational energy $E_{Q_0+q}$ of 
$|\Psi_{Q_0+q} \rangle$
\begin{equation}\label{varen}
\Delta(Q_0+q) = E_{Q_0+q} - E_{Q_0} = \frac{F_q^\alpha}
{\langle S_{-q}^\alpha S_{q}^\alpha \rangle},
\end{equation}
where $\langle \dots \rangle$ stands for the average value over 
$|\Psi_{Q_0} \rangle$ and the function $F_q^\alpha$ is given by the double 
commutator 
\begin{equation}\label{double0}
F^{\alpha}_q = \frac{1}{2} \langle \left[ S_{-q}^\alpha, 
\left[ {\cal H}, S_{q}^\alpha \right] \right] \rangle.
\end{equation}
By using standard commutation relations for the spin operators and by 
summing over all the spin components $\alpha$, we obtain
\begin{equation}\label{double}
\sum_{\alpha} F^{\alpha}_q = \frac{1}{2N} \sum_{k} 
\left[ J_{k+q} + J_{k-q} - 2J_{k} \right] 
\langle {\vec S}_{k} \cdot {\vec S}_{-k} \rangle.
\end{equation}
After performing the inverse Fourier transform, we have 
\begin{eqnarray}\label{disegua}
&& \sum_{\alpha}F^{\alpha}_q = 
\frac{1}{N} \sum_{i,j} J_{i,j} \{ \cos[q (R_i-R_j)] -1 \}
\langle {\vec S}_i \cdot {\vec S}_j \rangle \le \nonumber \\
&& \frac{q^2}{2N} \sum_{i,j} |J_{i,j}| |R_i-R_j|^2
|\langle {\vec S}_i \cdot {\vec S}_j \rangle| \le \tilde{J} S(S+1) q^2,
\end{eqnarray}
where we used that $|\langle {\vec S}_i \cdot {\vec S}_j \rangle| \le S(S+1)$ 
and the short-range condition~(\ref{shortrange}). Notice that the previous
bound does not rely upon a small-$q$ expansion, but it is instead valid for
all momenta.
Therefore, we arrive to the important result that:
\begin{equation}\label{final}
\Delta(Q_0+q) \le \frac{\tilde{J} S(S+1) q^2}{S(q)},
\end{equation}
where $S(q)=\langle S_{-q}^\alpha S_{q}^\alpha \rangle$ is the static magnetic
structure factor, that, in the SU(2) case, does not depend upon $\alpha$.
Since $|\Psi_{Q_0+q} \rangle$ is a variational state, its average energy has
to be higher than the lowest exact triplet excitation with given momentum
$Q_0+q$. We conclude, therefore,
that the inequality~(\ref{final}) will also hold for the exact triplet 
excitation energy with momentum $Q_0+q$. 

In the following we will argue that the rigorous bound on the excitation
spectrum~(\ref{final}) quite naturally implies the existence of a new branch 
of gapless excitations close to an hypothetical second-order phase transition
from a magnetic phase to a dimerized one. 
We will focus on the square lattice case and an 
ordered phase with staggered magnetization at $Q=(\pi,\pi)$, but similar
results can be also found in the case of different lattice structures or
spiral magnetic orders. 
For the sake of simplicity, we can imagine that $J_q$ depends upon one 
parameter $g$, proportional to the frustration of model: for instance, in the 
$J_1{-}J_2$ model $J_q = 2 (\cos q_x + \cos q_y) + 4 g \cos q_x \cos q_y$, 
with $g = J_2/J_1$. 
Suppose that, by increasing the frustrating parameter $g$, the system has a 
continuous transition from a magnetically ordered phase to a dimerized one. 
In this latter, the lowest triplet excitation remains gapped in the 
thermodynamic limit and the ground state is four-fold degenerate, with four 
different momenta $q=(0,0)$, $(0,\pi)$, $(\pi,0)$, and $(\pi,\pi)$. 
Again, the following arguments will also hold for any other dimer patterns, 
having ground states with momenta $q \ne (0,0)$ and $q \ne (\pi,\pi)$.
By applying the Feynman construction with $q=(\pi,\pi)$ to the state with
$q=(0,\pi)$ and by using the inequality~(\ref{final}), we obtain that
\begin{equation}\label{bound}
\Delta(\pi,0) \le \frac{2 \pi^2 \tilde{J} S(S+1)}{S(\pi,\pi)}.
\end{equation} 
In the dimerized region, i.e., for $g>g_c$, the spectrum is gapped, implying
that $S(\pi,\pi)$ is finite, and the constraint~(\ref{bound}) does not give
us any useful information on the excitations.
However, if the transition is continuous, the four-fold degeneracy will hold 
down to the critical point $g=g_c$, where $S(\pi,\pi)$ diverges, because of 
the incipient magnetic phase, stable for $g<g_c$. 
Indeed, at the critical point the spin-spin correlations are expected to 
decay as $|R|^{-(1+\eta)}$, leading to $S(\pi,\pi) \sim L^{1-\eta}$.
The actual value of $\eta$ can be found by considering the corresponding
classical model in three dimensions. In the Heisenberg model with 
nearest-neighbor interactions, we have that $\eta \sim 0$, implying a rather 
strong divergence of the static structure factor at the critical point. 
However, this divergence can be much weaker, i.e., $\eta \sim 0.5$, in the 
case where no free topological singularities are allowed but only pairs of 
them are present.~\cite{kamal,motrunich} 
We point out here that our conclusions remain valid as long as $\eta \le 1$, 
the structure factor being logarithmically divergent in the case of $\eta=1$.  

\begin{figure}
\includegraphics[width=0.45\textwidth]{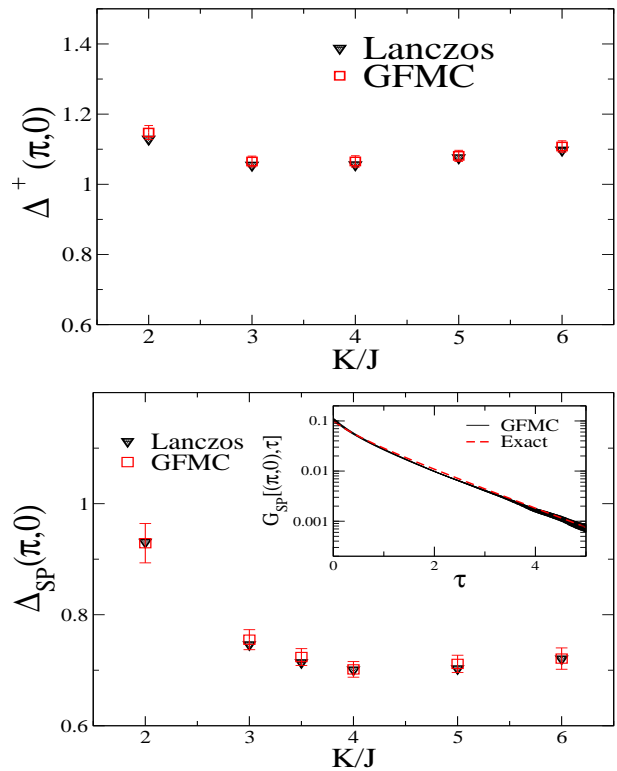}
\caption{\label{fig:conf}
(Color online) Comparison on a $6 \times 6$ lattice between Lanczos and GFMC 
for the lowest energy gap.
Upper panel: $\Delta^+(\pi,0)$ (see text) [in unit of $(K+J)$], as a 
function of $K/J$.
Lower panel: $\Delta_{SP}(\pi,0)$ (see text) [in unit of $(K+J)$], as 
a function of $K/J$. Inset: direct comparison of the dynamical correlation 
function for the dimer operator calculated by GFMC approach and by Lanczos.}
\end{figure}

Therefore, because the operator $S_q^\alpha$ used in the Feynman construction
carries $S=1$, we immediately arrive to the conclusion that 
Eq.~(\ref{bound}) implies the existence of a branch of excitations,
with $S=1$ and $q=(\pi,0)$, that becomes gapless at $g=g_c$. 
Analogously, by considering the ground state with momentum $q=(\pi,0)$, we
obtain a branch of excitations with $S=1$ and $q=(0,\pi)$.
This outcome contrasts the common understanding of the low-energy properties 
of magnetic systems with O(3) symmetry, which contains only two branches of 
gapless excitations at $q=(0,0)$ and $q=(\pi,\pi)$, while the 
triplet excitations at $q=(0,\pi)$ and $q=(0,\pi)$ are always 
gapped.~\cite{zinn}
Therefore, one is left with the following three possibilities: the first one
is to have an O(3) critical point with more than two gapless modes, 
not described by standard theories of magnetic phase transitions.
The second one is that the dimerized phase is separated from the magnetic 
phase by a first-order transition: in that case the spectrum can evolve
discontinuously and the triplet states at $q=(0,\pi)$ and $q=(0,\pi)$ can 
remain gapped. The third possibility is that there is still a continuous
transition but the non-magnetic phase does not break the translational 
invariance and, therefore, has no dimer order. Then, the dimerized phase 
could be eventually stabilized through a further (continuous) transition. 

\begin{figure}
\includegraphics[width=0.45\textwidth]{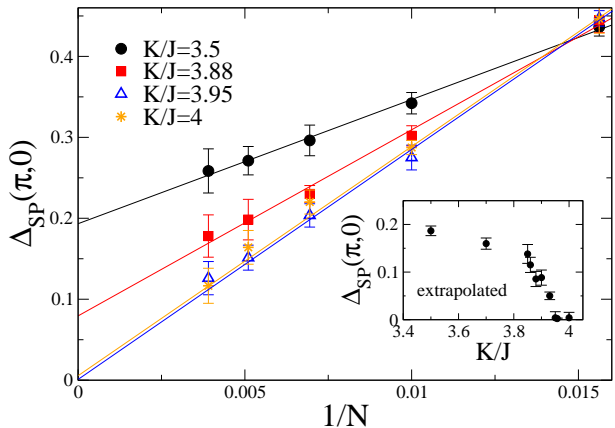}
\caption{\label{fig:scaldim}
(Color online) The dimer gap $\Delta_{SP}(\pi,0)$ [in unit of $(K+J)$] 
obtained by GFMC as a function of the size $N$ of the cluster for different 
values of $K/J$. 
The lines are linear fits of the points. Inset: the extrapolated value in 
the thermodynamic limit as a function of $K/J$.}
\end{figure}

We stress that a continuous transition between a magnetically ordered phase 
and a dimerized one is instead possible in models that {\it explicitly} 
break the translational invariance, like for instance 
the coupled ladder system,~\cite{ladder} where due to the folding of the 
Brillouin zone $q=(\pi,0)$ [or $q=(0,\pi)$] is equivalent to $q=(0,0)$.

Now we would like to discuss what happens for systems with different kinds 
of interactions. In the case of multi-spin interactions, it is easy to prove 
that we can arrive to similar expressions and the only difference is the 
prefactor, due to a different value of the double commutator~(\ref{double0}). 
A little more care must be paid to a model with U(1) symmetry. 
Here, the total spin $S$ does not
commute with the Hamiltonian, and only $S^z$ is a good quantum number.
Furthermore, the states with $S^z=0$ can be also classified according the 
discrete {\it particle-hole} symmetry ($PH$) $S^z \to -S^z$. 
In the weakly frustrated limit, without loss of generality, we can consider 
the ferromagnetic $XY$ model, that has only one gapless branch at $q=(0,0)$.
On a finite cluster, the ground state has $S^z=0$ and can be taken to be 
even under the PH symmetry, i.e., with $PH=+1$. In this case, we can construct
two different variational states for the low-energy excitations. 
The first one, with $S^z=1$, is given by 
$|\Psi_{Q_0+q} \rangle = S_q^+ |\Psi_{Q_0} \rangle$, $S_q^+$ being the
Fourier transform of $S_j^+=S_j^x+ i S_j^y$, and has
\begin{equation}\label{boundxy}
\Delta^+(Q_0+q) = \frac{F^+_q}{\langle S_{-q}^- S_{q}^+ \rangle}.
\end{equation}
Instead, the second possibility, with $S^z=0$ and $PH=-1$, is given by 
$|\Psi_{Q_0+q} \rangle = S_q^z |\Psi_{Q_0} \rangle$ and has
\begin{equation}\label{boundz}
\Delta^z(Q_0+q) = \frac{F^z_q}{\langle S_{-q}^z S_{q}^z \rangle}.
\end{equation}
In the dimerized phase the ground state is four-fold degenerate
with four different momenta $q=(0,0)$, $(0,\pi)$, $(\pi,0)$, and $(\pi,\pi)$, 
all of them having $S^z=0$ and $PH=+1$ [this is equivalent to have all 
singlets in the SU(2) case]. Then, in close relation with the SU(2) case, 
the Feynman construction~(\ref{boundxy}) with $q \to (0,0)$ applied to the 
state with momentum $q=(\pi,0)$ implies the existence of a gapless branch of 
excitations at $q=(\pi,0)$ with $S^z \ne 0$. 
Indeed, in this case $F^+_q$ is finite for $|q| \to 0$ and the structure 
factor in the denominator diverges at the critical point, as 
$\langle S_{-q}^- S_{q}^+ \rangle \propto 1/q^{1-\eta}$.
Hence, we arrive to the same conclusions as before: excluding the possibility
of having more than one gapless branch in a model with O(2) symmetry, 
in contrast with semi-classical approaches,~\cite{zinn} the transition 
must be either first order or second order to a non-dimerized phase.
Finally, it should be noted that, under an additional assumption, also 
Eq.~(\ref{boundz}) gives a strict bound to the spectrum.
Indeed, $F^z_q \propto q^2$ for $|q| \to 0$ and, whenever at the critical 
point $\langle S_{-q}^z S_{q}^z \rangle$ does not vanish too fast, i.e., 
$\langle S_{-q}^z S_{q}^z \rangle \propto q^{2-\alpha}$ with $\alpha > 0$, 
also $\Delta^z(q) \to 0$ for $|q| \to 0$.

\section{Numerical Results}\label{sec2}

In this section, we present our numerical results for the
Hamiltonian~(\ref{hamiring}), by using the Green's function Monte Carlo 
(GFMC) technique with a finite and fixed population of walkers.~\cite{gfmc}
In order to minimize the statistical fluctuations, we implemented the 
importance sampling, defined by the variational wave function:~\cite{franjic}
\begin{equation}\label{trial}
|\Phi_V \rangle = {\cal P}_z {\cal J} |F \rangle,
\end{equation}
where $|F \rangle$ is the ferromagnetic state with the magnetization along 
the $x$ direction, i.e., 
$|F \rangle = \Pi_{i}(|\uparrow \rangle_i +|\downarrow \rangle_i)$,
${\cal P}_z$ is the projector onto the subspace with $S^z=0$, 
and ${\cal J}$ is a spin Jastrow factor 
\begin{equation}
{\cal J} = {\rm exp} \left\{ \frac{1}{2} 
\sum_{i,j} v_{i,j} S^z_i S^z_j \right\},
\end{equation}
with the two-body potential $v_{i,j}$ that depends only upon the distance 
$|R_i-R_j|$ and can be optimized by using the minimization technique described
in Ref.~\onlinecite{sorella}.

\begin{figure}
\includegraphics[width=0.45\textwidth]{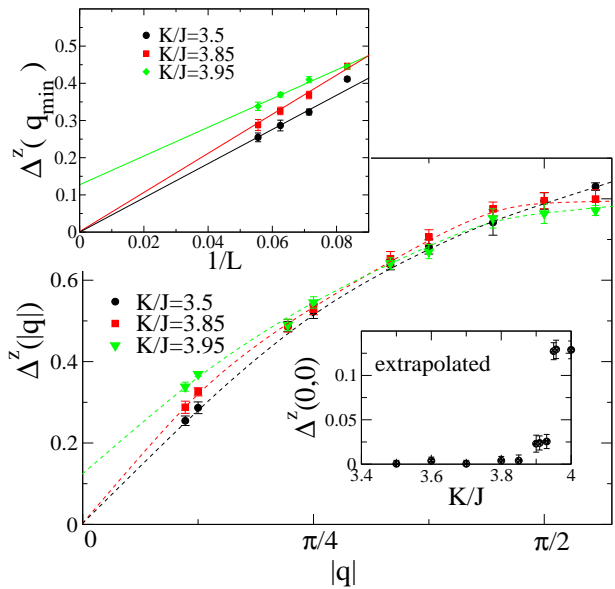}
\caption{\label{fig:spectrum}
(Color online) The spin-wave spectrum [in unit of $(K+J)$] obtained by GFMC 
by using the operator $O_q=S^z_q$. The linear sizes of the cluster are 
$L=16$ and $18$, and the lines are guides to the eye. The vector $q$ is 
taken along the $(1,0)$ direction. 
Upper inset: the behavior of the finite-size gap at the minimum momentum, 
i.e., $q_{\rm min}=(2\pi/L,0)$ as a function of $L$ for different values of 
$K/J$. 
Lower inset: the thermodynamic limit of the gap at $q=(0,0)$ as a function 
of $K/J$.}
\end{figure}

Since the Hamiltonian~(\ref{hamiring}) does not suffer from the sign 
problem, it is possible to sample the exact ground-state wave function
$\Phi_0(x)=\langle x|\Phi_0 \rangle$ (being $|x \rangle$ a generic spin 
configuration), or, with importance sampling, the positive quantity
$\Phi_0(x) \Phi_V(x) =\langle x|\Phi_0 \rangle \langle x|\Phi_V \rangle$.
Moreover, it is possible to evaluate the imaginary time evolution of the 
following dynamical correlation function:
\begin{equation}\label{green}
G(q,\tau) = 
\frac{\langle \Phi_V| O_{-q} e^{-\tau {\cal H}} O_{q} |\Phi_0 \rangle}
{\langle \Phi_V| e^{-\tau {\cal H}} |\Phi_0 \rangle},
\end{equation}
where $O_q$ is a given operator with a definite momentum $q$.
This technique has been already successfully used by one of us to compute the 
excitation spectrum of a quasi-one-dimensional magnetic system.~\cite{yunoki}
It is easy to show that, for large imaginary time $\tau$, $G(q,\tau)$ behaves
like:
\begin{equation}
G(q,\tau) \propto e^{-\tau \Delta_q},
\end{equation}
where $\Delta_q=E_q-E_0$ is the energy gap between the ground state and 
the first excited state $|\Phi_q \rangle$ such that 
\begin{equation}\label{nonort}
\langle \Phi_q|O_q|\Phi_0 \rangle \ne 0.
\end{equation}
Therefore, from the fit of the large-$\tau$ behavior of the dynamical 
correlation function, it is possible to extract the lowest gap with a given 
momentum $q$ and the symmetry properties imposed by the 
condition~(\ref{nonort}).

Whenever the operator $O_q$ is diagonal in the spin configuration $|x \rangle$,
like, for instance the case of $S_q^z$, the expression~(\ref{green}) can 
be easily calculated by using the standard forward walking 
technique.~\cite{gfmc} On the other hand, for non-diagonal operators, i.e.,
for $S_q^+$, a more involved calculation is needed, since the operator breaks
the Markov chain of the Monte Carlo simulation. In order to avoid this extra
numerical effort, we have calculated more efficiently $\Delta^+(q)$ by 
sampling the ground state with $S^z=1$ and by calculating its excitation 
spectrum, for instance by using $S_q^z$. Then, $\Delta^+(q)$ can be found from:
\begin{eqnarray}
\Delta^+(q) &=& E_q^{S^z=1}-E_0^{S^z=0} \nonumber \\
&=& (E_q^{S^z=1}-E_0^{S^z=1}) + E_0^{S^z=1} - E_0^{S^z=0},
\label{gapy}
\end{eqnarray}
where the term in the first bracket can be calculated by fitting $G(q,\tau)$
and the last two terms are just the ground-state energies of the two sectors 
with $S^z=0$ and $S^z=1$, easily calculated by GFMC.
 
In order to verify the accuracy of GFMC for the evaluation of the gap,
we can perform calculations on a $6 \times 6$ lattice, where the exact 
results are available by the Lanczos method. In particular, we consider 
the dimer operator 
$O_q = O^{SP}_q = 1/\sqrt{N} \sum_j e^{i q R_j} S_j^z S_{j+x}^z$
with $q=(\pi,0)$. In this case, the operator $O_q$ does not change the $PH$
quantum number and we have access to the lowest gap with the same $PH$ of 
the ground state.
In Fig.~\ref{fig:conf}, we compare, for different values of the ring-exchange 
coupling $K/J$, the exact energy gap at $q=(\pi,0)$ and even under $PH$ with 
the one extracted from the large-$\tau$ behavior of $G(q,\tau)$, and denoted 
by $\Delta_{SP}(\pi,0)$.
Moreover, as a further check of the statistical approach, we can also compute 
exactly the dynamical correlation function and make the comparison with 
the GFMC calculations (see the inset of Fig.~\ref{fig:conf}). It should be 
stressed that, for our purpose, we are only interested in the lowest energy 
gap, and, therefore, we fit only the large-$\tau$ part of $G(q,\tau)$. 
This can be achieved by using only two or three exponents at 
most.~\cite{maxent} 
In the same figure, we also report the comparison for $\Delta^+(\pi,0)$,
extracted from the decomposition of Eq.~(\ref{gapy}): also in this case
the agreement is excellent.

\begin{figure}
\includegraphics[width=0.45\textwidth]{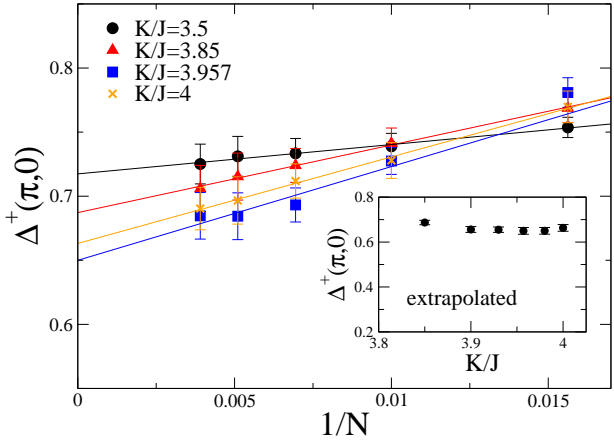}
\caption{\label{fig:main}
(Color online) The gap $\Delta^+(\pi,0)$ [in unit of $(K+J)$] calculated by 
GFMC as a function of the size $N$ of the cluster for different 
values of $K/J$. The lines are linear fits of the points. Inset: the 
thermodynamic value of the gap as a function of $K/J$.}
\end{figure}

The lowest dimer gap $\Delta_{SP}(\pi,0)$ provides us the informations on 
the appearance of a dimerized phase with spin-Peierls order. 
In Fig.~\ref{fig:scaldim}, we report the size-scaling calculations of
$\Delta_{SP}(\pi,0)$ as a function of the inverse of the size $N$, for 
different values of the ring exchange. The dimer gap is clearly finite in the 
magnetically ordered phase, with a decreasing behavior by approaching the 
transition. The thermodynamic limit indicates a vanishing value 
of $\Delta_{SP}(\pi,0)$ for $K/J \gtrsim 3.95$, in close agreement
with Ref.~\onlinecite{sandvik}.
 
Let us now turn to the spin-wave operators $S_q^+$ and $S_q^z$, in order to
apply the arguments developed in the previous section and assess the nature
of the transition. First of all, it should be noticed that, in the $XY$ 
model, $S_q^z$ acts on the ground state by creating a standard spin-wave 
excitation. This can be easily seen by considering the Holstein-Primakoff 
representation at the leading order in the $1/S$ expansion:
\begin{eqnarray}
S_j^x &=& S-a_j^{\dagger}a_j \nonumber \\
S_j^y &=& \sqrt{\frac{S}{2}}(a_j^{\dagger}+a_j) \nonumber \\
S_j^z &=& -i\sqrt{\frac{S}{2}}(a_j^{\dagger}-a_j). \nonumber
\end{eqnarray}
After standard calculations, we arrive to the result that
\begin{eqnarray}
S_q^y |0 \rangle &\sim& \alpha_q^{\dagger} |0 \rangle \nonumber \\
S_q^z |0 \rangle &\sim& \alpha_q^{\dagger} |0 \rangle, \nonumber
\end{eqnarray}
where $|0 \rangle$ is the spin-wave ground state and $\alpha_q^{\dagger}$ is 
the creation operator of the elementary excitation, defined by the Bogoliubov
transformation.
Therefore, the states of the gapless branch have non-zero overlap with the
(normalized) states generated either by $S_q^y$ or by $S_q^z$, allowing us
to assess directly the spin-wave spectrum.
As shown in Fig~(\ref{fig:spectrum}), for small values of the ring-exchange
coupling, the excitation spectrum remains gapless for $|q| \to 0$ and the 
opening of a gap, for $K/J \sim 3.95$, signals the transition to the disordered 
phase. Moreover, in the ordered phase, the spectrum is linear for small 
momenta, i.e., $\Delta^z(q) \sim c|q|$ (see Fig.~\ref{fig:spectrum}), 
the constant $c$ defining the spin-wave velocity. 
From our numerical results, it comes out that, as assumed in semi-classical 
approaches, the spin velocity $c$ is not strongly renormalized by frustration
and remains finite up to the transition.

\begin{figure}
\includegraphics[width=0.45\textwidth]{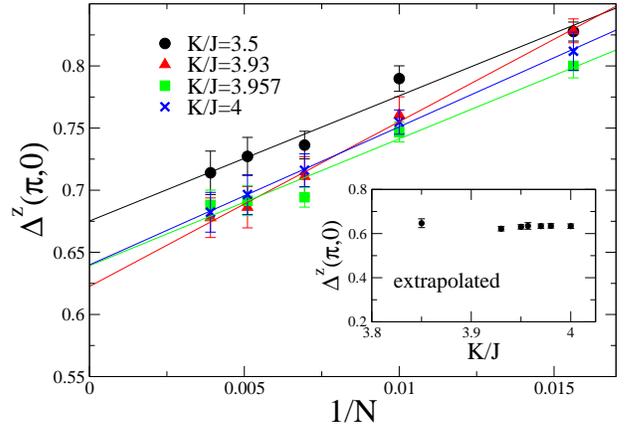}
\caption{\label{fig:main2}
(Color online) The gap $\Delta^z(\pi,0)$ [in unit of $(K+J)$] calculated by 
GFMC as a function of the size $N$ of the cluster for different values of 
$K/J$. The lines are linear fits of the points. 
Inset: the thermodynamic value of the gap as a function of $K/J$.}
\end{figure}

By using the same spin-wave operators $S_q^+$ and $S_q^z$ at $q=(\pi,0)$ and
$q=(0,\pi)$, we can assess the nature of the transition between the 
ferromagnet and the dimer state.
In particular, according to Eq.~(\ref{boundxy}), a continuous phase transition
implies a vanishing $\Delta^+(q)$ for $q=(\pi,0)$ and $q=(0,\pi)$.
In Fig.~(\ref{fig:main}), we report the results for the gap 
$\Delta^+(\pi,0)$ for different ratios $K/J$ and various sizes of the cluster.
From our results, it turns out rather clearly that $\Delta^+(\pi,0)$ remains 
finite in the thermodynamic limit for all the values of the ring-exchange 
couplings, also very close to the transition point.
Of course, we cannot rule out the existence of a very tiny region, near to
the transition point, in which this gap would drop down to zero.
Another possibility for missing the vanishing behavior of the gap could be
due to an anomalous size scaling, that changes its behavior beyond a
characteristic length scale $L_c$ much bigger than the values of $L$ 
considered here. 
However, from our GFMC calculations, we do not have any sizable sign that
could indicate the existence of gapless excitation at $q=(\pi,0)$ or 
$q=(0,\pi)$, and even at the transition point $K_c/J \sim 3.957$ the 
extrapolated value of $\Delta^+(\pi,0)$ remains clearly finite.
Since a clear dimer order appears as soon as we enter the disordered 
region,~\cite{sandvik} we can safely rule out the possibility to have a 
spin-liquid phase in between the ferromagnetic and the dimerized phases. 
Therefore, our results on
$\Delta^+(\pi,0)$ strongly suggest that a first-order transition is the more 
likely scenario in this O(2) model with ring-exchange interactions.
Similarly, also the gap $\Delta^z(q)$ at $q=(\pi,0)$ and $q=(0,\pi)$ does not
show any evidence of softening and remains finite at the transition, see 
Fig.~\ref{fig:main2}.

\section{Conclusions}\label{sec3}

In conclusion, we investigated the excitation spectrum of a spin model close
to the transition from a magnetic phase to a dimerized one, which is 
particularly important in view of the theoretical predictions of Senthil and 
coworkers. In particular, we considered an $XY$ model in presence of a 
frustrating ring-exchange interaction, a model that certainly leads to a
magnetic-dimer transition, as recently shown by Sandvik and 
collaborators.~\cite{sandvik,sandvik2}

The central part of our work is the derivation of a general constraint for the
excitation spectrum for a broad class of short-range spin Hamiltonians: 
whenever the transition is second-order a new branch of gapless excitations 
must exists, besides the ones expected from standard semi-classical approaches. 
In particular, on the square lattice with a typical four-fold dimer ground 
state, these modes are spin-wave-like excitations at momenta $q=(0,\pi)$ 
and $q=(\pi,0)$. On the contrary, by using a numerically exact 
technique, we did not find any evidence of these gapless modes.
Therefore, our results rule out both conventional and unconventional 
continuous transition, implying that a first-order transition represents the 
most likely scenario, in agreement with recent calculations.~\cite{note2}
Although we cannot exclude anomalous finite-size effects, that 
should be particularly important for correlation functions 
(see Ref.~\onlinecite{sandvik}), we expect that the excitation spectrum 
does not suffer from these problems, depending only on total 
energy differences.

We acknowledge the constant and fruitful interaction with A. Parola, 
A. Sandvik, and R. Melko. We tank especially L. Capriotti who was involved
in the early stage of this project. F.B. thanks F. Alet for very useful 
discussions. This research has been supported by PRIN-COFIN 2004 and INFM.
F.B. acknowledges the warm hospitality in the University of Toulouse
and its partial support.


\begin{thebibliography}{99}

\bibitem{diep} For a recent review, see for instance,
   {\it Frustrated Spin systems}, Edited by H.T. Diep (World Scientific, 
   Singapore, 2003).

\bibitem{fazekas} P. Fazekas and P.W. Anderson, Philos. Mag. {\bf 30}, 423 
   (1974).

\bibitem{coldea} R. Coldea, D.A. Tennant, A.M. Tsvelik, and Z. Tylczynski,
   \prl {\bf 86}, 1335 (2001); R. Coldea, D.A. Tennant, and Z. Tylczynski,
   \prb {\bf 68}, 134424 (2003).

\bibitem{kanoda} Y. Shimizu, K. Miyagawa, K. Kanoda, M. Maesato, and G. Saito,
   \prl {\bf 91}, 107001 (2003).

\bibitem{kagome} A.P. Ramirez, G.P. Espinosa, and A.S. Cooper, \prl {\bf 64},
   2070 (1990).

\bibitem{oned} See e.g. the review on ${\rm CuGeO_3}$ by J.-P. Boucher and
   L.-P. Regnault, J. Physique {\bf 6}, 1939 (1996).

\bibitem{wen} X.-G. Wen, \prb {\bf 65}, 165113 (2002).

\bibitem{hastings} M.B. Hastings, \prb {\bf 69}, 104431 (2004).

\bibitem{senthil1} T. Senthil, A. Vishwanath, L. Balents, S. Sachdev, and 
   M.P.A. Fisher, Science {\bf 303}, 1490 (2004).

\bibitem{senthil2} T. Senthil, L. Balents, S. Sachdev, A. Vishwanath, 
   and M.P.A. Fisher, \prb {\bf 70}, 144407 (2004).

\bibitem{sandvik} A.W. Sandvik, S. Daul, R.R.P. Singh, and D.J. Scalapino,
   \prl {\bf 89}, 247201 (2002). 

\bibitem{balents} A similar model with a ring-exchange term on the Kagome 
   lattice in the Ising-axis limit has been predicted to exhibit a non-trivial
   quantum phase transition from a magnetically ordered to a spin-liquid phase.
   D.N. Sheng and L. Balents, \prl {\bf 94}, 146805 (2005).

\bibitem{sandvik2} R.G. Melko, A.W. Sandvik, and D.J. Scalapino, \prb {\bf 69},
   100408 (2004).

\bibitem{thanks} We thanks R.G. Melko for providing us the best estimate of 
   the transition point. 

\bibitem{note2} From recent numerical calculations, it seems that, from the
   disordered region, the dimer order parameter vanishes at the transition 
   point, while, from the ordered phase, the spin stiffness shows a small 
   jump. R. Melko and A.W. Sandvik, private communication.

\bibitem{gfmc} M. Calandra Buonaura and S. Sorella, \prb {\bf 57}, 
   11446 (1998).

\bibitem{feynman} R.P. Feynman and M. Cohen, Phys. Rev. {\bf 102}, 1189 (1956).

\bibitem{kamal} M. Kamal and G. Murthy, \prl {\bf 71}, 1991 (1993). 

\bibitem{motrunich} O.I. Motrunich and A. Vishwanath, \prb {\bf 70}, 075104
   (2004). 

\bibitem{zinn} For a review on the semi-classical approaches and their
   relation with field-theory methods, see for instance, J. Zinn-Justin,
   {\it Quantum Field Theory and Critical Phenomena} (Clarendon Press,
   Oxford, 1989).

\bibitem{ladder} L. Capriotti and F. Becca, \prb {\bf 65}, 092406 (2002).

\bibitem{franjic} F. Franjic and S. Sorella, Prog. Teor. Phys. {\bf 97}, 
   399 (1997)

\bibitem{sorella} S. Sorella, \prb {\bf 64}, 024512 (2001).

\bibitem{yunoki} S. Yunoki and S. Sorella, \prl {\bf 92}, 157003 (2004).

\bibitem{maxent} In this respect, more sophisticated techniques are possible,
   like for instance the maximum entropy method, that, however, do not change 
   the estimate of the lowest gap excitation.

\end{thebibliography}
\end{document}